\documentclass{article}
\pdfoutput=1
\PassOptionsToPackage{numbers, sort, compress}{natbib}
\usepackage[final]{neurips_2020}

\usepackage{epsfig,endnotes}
\usepackage{varwidth}
\usepackage{algorithm}
\usepackage{algpseudocode}
\usepackage{balance}
\usepackage{xcolor}
\usepackage{nicefrac}
\usepackage{amsmath}
\usepackage{braket}
\usepackage{bm}
\usepackage{mathtools}
\usepackage{multirow}
\usepackage{bigdelim}
\usepackage{mathtools}
\usepackage{amssymb}
\usepackage{amsfonts}
\usepackage{algorithmicx}
\usepackage{indentfirst}
\usepackage{booktabs}

\usepackage{enumitem}
\usepackage{boxedminipage}
\usepackage{float}
\usepackage{graphicx}
\usepackage{caption}
\usepackage{subcaption}
\usepackage[normalem]{ulem}
\usepackage{xpatch}
\usepackage{xspace}
\usepackage{listings}
\usepackage{mathpartir}
\usepackage{makecell}
\usepackage[hyperfootnotes=false]{hyperref}%
\usepackage{cleveref}
\usepackage{textcomp}
\usepackage{framed,mdframed}
\usepackage{tablefootnote}

\usepackage{flushend}

\usepackage{verbatim}
\usepackage{fancyvrb}
\usepackage{amsthm}

\usepackage{stmaryrd}

\usepackage{tablefootnote}

\newlength{\saveparindent}
\newlength{\saveparskip}
\newcommand{\tensorflow}{{{TensorFlow}}\xspace}
\newcommand{\tool}{{\textsc{CrypTFlow}}\xspace}

\newcommand{\mpc}{{MPC}\xspace}

\newcommand{\namedref}[2]{\hyperref[#2]{#1~\ref*{#2}}\xspace}

\definecolor{mypink}{rgb}{1,0.2,0.4}

\newcommand{\bbZ}{\mathbb{Z}}

\newcommand{\cryptflow}{\textsc{CrypTFlow}}
\newcommand{\resnet}{{\textsc{ResNet50}}\xspace}

\newcommand{\densenet}{{\textsc{DenseNet121}}\xspace}

\usepackage{url}

\usepackage[utf8]{inputenc} %
\usepackage[T1]{fontenc}    %
\usepackage{hyperref}       %
\usepackage{url}            %
\usepackage{booktabs}       %
\usepackage{amsfonts}       %
\usepackage{nicefrac}       %
\usepackage{microtype}      %

\title{Secure Medical Image Analysis with CrypTFlow}

\author{%
  Javier Alvarez-Valle\\
  MSR UK
  \And
  Pratik Bhatu\\
  MSR India
  \And
  Nishanth Chandran\\
  MSR India
  \And
  Divya Gupta\\
  MSR India
  \And
  Aditya Nori\\
  MSR UK
  \And
  Aseem Rastogi\\
  MSR India
  \And  
  Mayank Rathee\\
  MSR India
  \And
  Rahul Sharma\\
  MSR India
  \And
  Shubham Ugare\\
  UIUC, USA
}

\begin{document}

\maketitle

\begin{abstract}
We present \cryptflow, a system that converts
\tensorflow inference code into Secure
Multi-party Computation (MPC) protocols at the push of a
button. To do this, we build two components. Our first component  is an end-to-end compiler from \tensorflow
to a variety of MPC protocols. The second
component is an improved semi-honest 3-party protocol
that provides significant speedups for inference.
We empirically demonstrate
the power of our system by showing the secure inference of real-world
neural networks such as \densenet\ for detection of lung diseases from chest X-ray images
and 3D-UNet for segmentation in radiotherapy planning using CT images. In particular, this paper provides the first evaluation of secure {\em segmentation} of 3D images, a task that requires much more powerful models than classification and is the largest secure inference task run till date.
\end{abstract}

\section{Introduction}
Secure multiparty computation (MPC) allows a set of mutually distrusting parties to
compute a publicly known function on their secret inputs without revealing their
inputs to each other. This is done through the execution of a cryptographic protocol which guarantees that the protocol participants learn only the function output on their secret inputs and nothing else. 
\mpc has made rapid strides - from
being a theoretical concept three decades ago \cite{yao,gmw}, to now
being on the threshold of having real world impact.
One of the most compelling use cases for MPC is machine
learning (ML) - e.g. being able to do secure ML inference when the model and the query are private inputs belonging to different parties.
There has been a flurry of
recent works aimed at running inference securely with MPC such as
SecureML~\cite{secureml}, MinioNN~\cite{minionn}, 
ABY$^3$~\cite{aby3}, CHET~\cite{chet},
SecureNN~\cite{securenn}, Gazelle~\cite{gazelle}, Delphi~\cite{delphi}, and so on.
 Unfortunately, these techniques are not easy-to-use by ML developers and have only been demonstrated on small deep
 neural networks (DNNs) on tiny datasets such as MNIST and CIFAR.
However, in order for MPC to be truly ubiquitous for secure inference
tasks, it must be both easy to use by developers with no  background in cryptography and capable of scaling to the  DNNs used in practice. 

In this work, we present \cryptflow, a  system that
converts \tensorflow \cite{tensorflow} inference code into \mpc protocols at the push of a button. By converting code in
standard \tensorflow, a ubiquitous ML framework that is used in production by
various technology companies, to \mpc protocols, \cryptflow\ significantly lower the entry barrier for ML practitioners and
programmers to use cryptographic \mpc protocols in real world
applications. We make the following  contributions:

 First, for the developer frontend, we provide a  compiler, called {\em Athos}, from
  \tensorflow to a variety of secure computation protocols (both 2 and
  3 party) while preserving accuracy. 
  The compiler is designed to be modular  and it provides facilities for plugging in different \mpc protocols.
To demonstrate this modularity, we have integrated Athos with the following backends: ABY-based~\cite{aby}  2-party computation (2PC), 
SCI-based 2PC~\cite{cryptflow2},
Aramis-based  malicious secure 3-party computation~\cite{cryptflow}, and Porthos-based semi-honest secure 3-party computation (3PC). 

 Second, for the cryptographic backend, we provide a  semi-honest secure 3-party computation protocol, {\em Porthos}, that outperforms all prior protocols for secure inference and enables us to execute, 
for the first time, cryptographically secure inference of ImageNet scale networks. Prior work in the area of secure inference has been limited to small networks over tiny datasets such as MNIST or CIFAR.
We have evaluated \tool on secure inference over DNNs that are at least an order of magnitude larger
than the state-of-the-art~\cite{delphi,chet,chameleon,securenn,secureml,gazelle,ezpc,minionn,aby3,nhe,xonn,nitin}.
Even on MNIST/CIFAR, Porthos has lower communication complexity and is more efficient than prior works~\cite{securenn,aby3,chameleon}. 

 Third, we demonstrate the ease-of-use, efficiency and scalability of \cryptflow\ by evaluating on \resnet~\cite{resnet} for ImageNet classification, \densenet~\cite{densenet} for detection of lung diseases from chest X-ray images and 3D-UNet~\cite{unet} for segmentation of raw 3D CT images.

Our toolchain is publicly
available\footnote{\url{https://github.com/mpc-msri/EzPC}}.
This paper reviews  the original \tool paper~\cite{cryptflow} briefly and its increment lies in the secure segmentation evaluation (Section~\ref{sec:unet}).

\section{Athos}
 Athos compiles \tensorflow inference code to secure computation protocols. 
The transformations implemented in Athos are sensitive to the performance of \mpc protocols. 
 For performance reasons all efficient secure computation protocols perform computation over fixed-point arithmetic - i.e., arithmetic over integers or arithmetic with fixed precision. 
 Athos automatically converts \tensorflow code over floating-point values into code that computes the same function over fixed-point values. This compilation is done while {\em matching} the inference accuracy of floating-point code. 
Prior works (\cite{secureml,minionn,gazelle,aby3,securenn,delphi}) in the area of running ML securely have performed this task by hand with significant losses in accuracy over floating-point code.

Athos represents a 32-bit floating-point number $r$ by a 64-bit integer $\lfloor r.2^s\rfloor$ for a precision or scale $s$. Then operations on 32-bit floating-point numbers are simulated by operations on 64-bit integers. For example $r_1\times r2$ is simulated as $\frac{\lfloor r_1.2^s\rfloor\times\lfloor r_1.2^s\rfloor}{2^{s}}$.
A large $s$ causes integer overflows and a small $s$ leads to accuracy loss.
To obtain a suitable scale $s$ (all variables have the same precision in Athos output),
 Athos works by ``sweeping through'' various precision levels to estimate the best precision~\cite{cryptflow}. 

\section{Porthos}
Porthos is an improved semi-honest 3-party secure computation protocol (tolerating one corruption) that builds upon SecureNN~\cite{securenn}. 
Porthos makes two crucial modifications to SecureNN. 
First, SecureNN reduces convolutions  to matrix multiplications and  invokes the Beaver triples~\cite{beaver} based matrix multiplication protocol. 
When performing a convolution with filter size $f\times f$ on a matrix of size $m\times m$, the communication is roughly $2q^2f^2+2f^2+q^2$ elements in the ring $\bbZ_{2^{64}}$, where $q = m-f+1$. 
Porthos computes these Beaver triples by appropriately reshaping $m\times m$ and $f\times f$ matrices. 
This reduces the communication to roughly $2m^2+2f^2+q^2$ ring elements. 
Typically the filter size, $f$, is between 1 and 11 and the communication of Porthos can be up to two orders of magnitudes less than SecureNN. 
Additionally, in SecureNN, the protocols for non-linear layers (such as ReLU and MaxPool)  require the third party to send secret shares to the first two parties. 
In Porthos, we cut this communication to half by eliminating the communication of one of these shares~\cite{cryptflow}. 
This reduces the communication in the overall ReLU and MaxPool protocols by 25\%. 

\section{Motivating Example}\label{sec:toolchain}

\begin{figure}
\centering
\begin{minipage}{.7\textwidth}
  \centering
  \includegraphics[width=\linewidth]{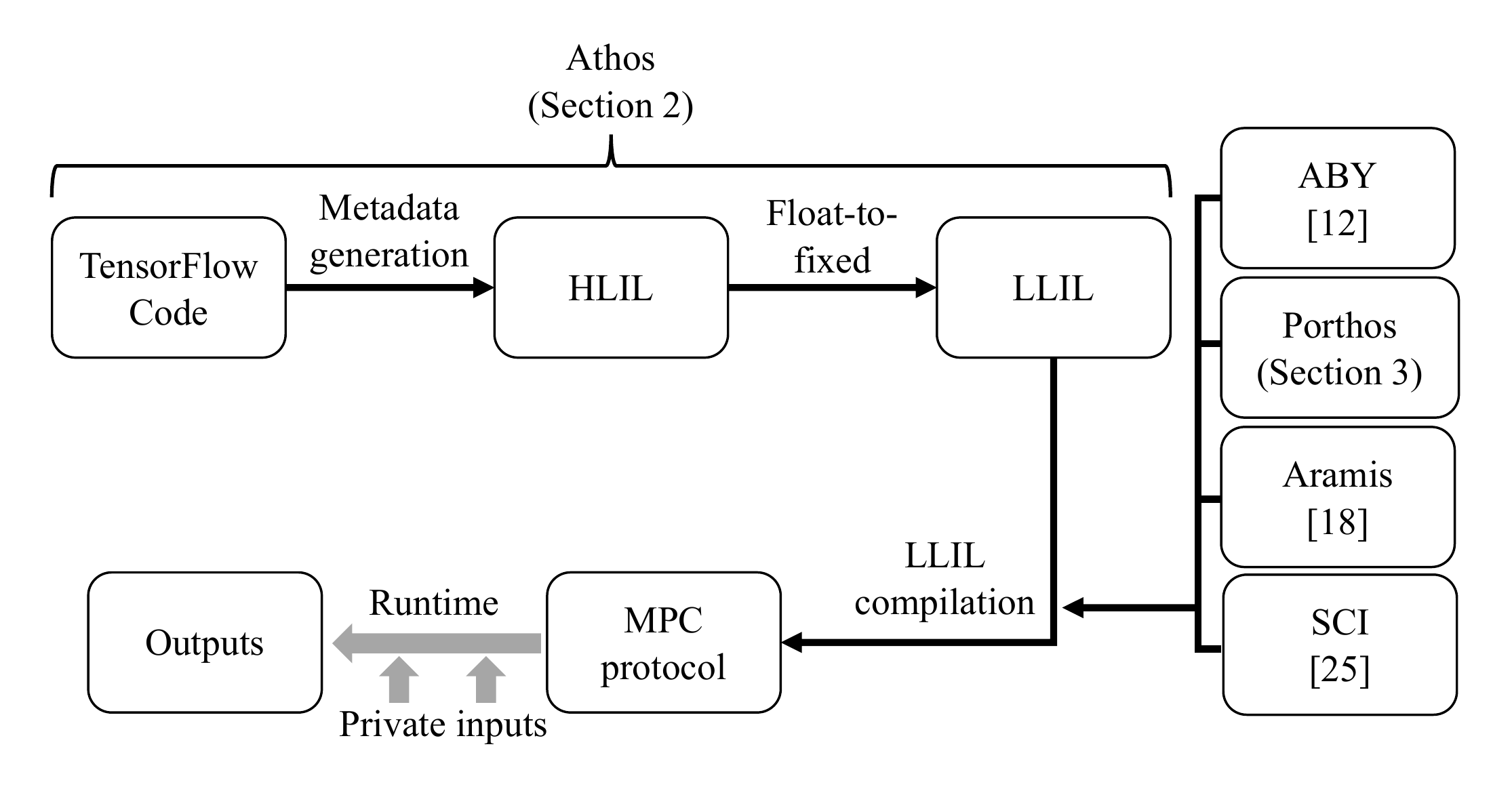}
  \caption{\cryptflow: End-to-end toolchain}
  \label{fig:cryptflowtoolchain}
\end{minipage}%
\begin{minipage}{.3\textwidth}
  \centering\small
\begin{Verbatim}
# x is (1,784) MNIST image.
# W and b are model parameters.

print(tf.argmax(tf.matmul(x, W) + b, 1))
\end{Verbatim}
\caption{Logistic Regression: TensorFlow snippet}
\label{fig:lrtf}
\end{minipage}
\end{figure}

In this section, we describe the end-to-end working of \cryptflow\ through an example of logistic regression. The toolchain is shown in Figure \ref{fig:cryptflowtoolchain}. 

\begin{figure}
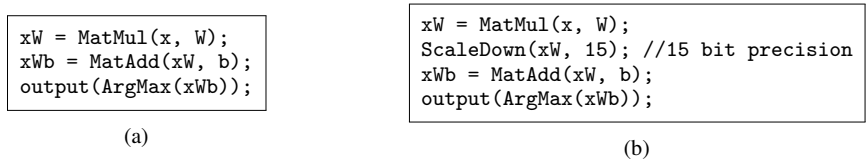

  \centering
  \resizebox{0.47\columnwidth}{!}{
    \begin{subfigure}{0.49\columnwidth}
      \centering
      \setlength{\fboxsep}{1.7mm}
      \fbox{\BUseVerbatim[fontsize=\small]{HLIL_LR_Verbatim}}
      \caption{}
      \label{fig:lrseedot}
    \end{subfigure}
  }
  \resizebox{0.47\columnwidth}{!}{
    \begin{subfigure}{0.49\columnwidth}
      \centering
      \setlength{\fboxsep}{1.6mm}
      \fbox{\BUseVerbatim[fontsize=\small]{LLIL_LR_Verbatim}}
      \caption{}
      \label{fig:lrezpc}
    \end{subfigure}
  }
  \caption{Logistic Regression in (a) floating-point: HLIL syntax (b) fixed-point: LLIL syntax}
\end{figure}

\cryptflow\  takes as input a pre-trained floating-point
\tensorflow model. For example, consider the code snippet for
logistic regression over MNIST
dataset in \tensorflow as shown in Figure \ref{fig:lrtf}. 
Our compiler
 first generates the
\tensorflow graph dump  as
well as metadata containing the dimensions of all the tensors.
Next, the \tensorflow graph
dump is compiled into a high-level intermediate language HLIL. The
code snippet for logistic regression in HLIL is shown in Figure
\ref{fig:lrseedot}. Next, Athos' float-to-fixed converter translates
the floating-point HLIL code to fixed-point code in a low-level
intermediate language LLIL. This step requires Athos to
compute the right precision to be used for maximum accuracy
Figure \ref{fig:lrezpc} shows the
LLIL code snippet for logistic regression. The operation $\mathtt{ScaleDown(X,s)}$ divides each 64-bit integer entry of tensor $X$ by $2^s$.
The function calls in the LLIL
code can be implemented with a variety of secure computation
backends - e.g. ABY~\cite{aby} for the case of 2-party secure
computation, Porthos for the case of semi-honest 3-party secure
computation, and Aramis~\cite{cryptflow} for the malicious secure variant. Different backends
provide different security guarantees and hence vary in their
performance. For this example, the three backends take
227ms, 6.5ms, and 10.2ms respectively.

\begin{SaveVerbatim}{HLIL_LR_Verbatim}
xW = MatMul(x, W);
xWb = MatAdd(xW, b);
output(ArgMax(xWb));
\end{SaveVerbatim}

\begin{SaveVerbatim}[]{LLIL_LR_Verbatim}
xW = MatMul(x, W);
ScaleDown(xW, 15); //15 bit precision
xWb = MatAdd(xW, b);
output(ArgMax(xWb));
\end{SaveVerbatim}

\section{Experiments}\label{sec:experiments}

\noindent\textbf{Overview.} First, in Section \ref{subsec:bigbenchmarks}, we use \tool for secure classification on ImageNet  using the following pre-trained \tensorflow models: \resnet\footnote{\url{https://github.com/tensorflow/models/tree/master/official/r1/resnet}} and \densenet\footnote{\url{https://github.com/pudae/tensorflow-densenet}}. 
We show that the fixed-point MPC protocols generated by Athos matches the accuracy of cleartext floating-point  \resnet and \densenet.  We also show how the optimizations in Porthos help it outperform prior works in terms of communication complexity and overall execution time. Finally, we discuss two case-studies of running \cryptflow\ on DNNs for medical image analysis.
The compilation time of \tool is around 5 sec for \resnet, 35 sec for \densenet and 2 minutes for 3D UNet.
\subsection{Secure Inference on ImageNet}\label{subsec:bigbenchmarks}
These experiments are in a LAN setting on 3.7GHz machines, each with 4 cores and with 16 GB of RAM. The measured bandwidth between each of the machines was at most 377 MBps and the latency was sub-millisecond. 

\resnet takes 25.9 seconds and 6.9 GB of communication; \densenet takes 36 seconds and 10.5 GB of communication.
 We measure communication as total communication between all $3$ parties - each party roughly communicates a third of this value. 
We show that Athos generated fixed-point code matches the accuracy of floating-code on \resnet and \densenet in Table~\ref{tab:fixed-accuracy}.

\begin{table}
\parbox{.45\linewidth}{
\centering
\begin{tabular}{|c|c|c|c|c|}
\hline
Benchmark & Float & Fixed & Float & Fixed  \\
          & Top 1 & Top 1 & Top 5 & Top 5  \\
\hline
$\resnet$     & 76.47 & 76.45 & 93.21 & 93.23  \\ \hline
$\densenet$   & 74.25 & 74.33 & 91.88 & 91.90  \\ \hline
\end{tabular}
\caption{Accuracy of fixed- vs floating-point.}
\label{tab:fixed-accuracy}
}
\hfill
\parbox{.45\linewidth}{
\centering
     \begin{tabular}{|c|c|c|c|}
    \hline
    SecureNN & Porthos& SecureNN & Porthos \\
      (s)  & (s) & Comm. (GB) & Comm. (GB) \\
	\hline
	 $38.36$ & $25.87$& $8.54$& $6.87$ \\
	\hline
     $53.99$ & $36.00$& $13.53$ & $10.54$ \\ 	\hline

\end{tabular}
 \caption{Porthos vs SecureNN.}
\label{tab:porthosvssecurenn}
}
\end{table}

Detailed comparisons of \tool with prior works on secure inference can be found in~\cite{cryptflow}.
However, since Porthos builds on SecureNN, we compare them  on ImageNet scale benchmarks in Table \ref{tab:porthosvssecurenn}. For this purpose, we add the code of SecureNN available at~\cite{securenncode} as another backend to \tool. These results show that Porthos improves upon the communication of SecureNN by a factor of roughly 1.2X--1.5X and the runtime by a factor of roughly 1.4X--1.5X.

\subsection{Lung diseases from 2D chest X-Ray images} 
In \cite{chestxray2018}, the authors train a {\densenet} to predict lung diseases from chest X-ray images. They use the publicly available NIH dataset of chest X-ray images and end up achieving an average AUROC score of 0.845 across 14 possible disease labels. These DNNs are available as pre-trained Keras models. We converted them into \tensorflow using~\cite{kttf} and compiled the automatically generated \tensorflow code with \tool.
During secure inference, we observed no loss in classification accuracy and the latency is similar to the runtime of \densenet for ImageNet.

\subsection{Segmenting tumors and organs at risk from 3D CT images}
\label{sec:unet}
Half a million cancer patients receive radiotherapy each year~\cite{demand}. Personalized radiation treatments require segmenting tumors and organs at risk from 3D volumetric images. Currently, this segmentation is a manual process where an oncologist draws contours along regions of interest slice-by-slice across the whole volume. This process often takes several hours per image which ML provided automation~\cite{stan,shuai} can reduce to minutes. We consider a 3D-UNet model~\cite{unet} that takes as input a raw 3D image obtained via Computed Tomography (CT) scans of the pelvic region and delineates tumor volumes and organs at risk.
This model's accuracy is within the inter-observer variability seen among clinical experts~\cite{innereye} and requires 1.87 Teraflops per inference.

Since this model is implemented in PyTorch, we first export it to ONNX and then use \tool's ONNX frontend. 
 For our secure inference setup,  each party has 32 cores running at 2.4GHz, no GPUs, and 128GB RAM. The parties are connected on a network with ping latency 0.2s and 625MBps bandwidth. On this set up, secure inference incurs a latency of 1 hour and 57 minutes and 557GB of communication. The most expensive operators in this computation are 3D transposed convolutions (or deconvolutions) and, to the best of our knowledge, \tool is the only secure inference tool that supports these operations.
 In our experience, it takes a couple of days for a scan to reach the oncologist for review and hence this latency overhead can be acceptable. 

\section{Related work and conclusion}
Other related systems
for converting PyTorch/Tensorflow to MPC protocols~\cite{crypten,tfe,pysyft,quantizednn} only support  3PC. Whereas, \tool additionally supports 2PC backends.
\tool provides the first implementation and evaluation of a system for secure segmentation.
With \tool, data scientists, with no background in cryptography, can obtain secure inference implementations for their trained models at the push of a button.

\begingroup
  \small
  \bibliography{main}

\begin{thebibliography}{33}
\providecommand{\natexlab}[1]{#1}
\providecommand{\url}[1]{\texttt{#1}}
\expandafter\ifx\csname urlstyle\endcsname\relax
  \providecommand{\doi}[1]{doi: #1}\else
  \providecommand{\doi}{doi: \begingroup \urlstyle{rm}\Url}\fi

\bibitem[cry(2019)]{crypten}
{Crypten by Facebook}, 2019.
\newblock URL \url{https://github.com/facebookresearch/CrypTen}.

\bibitem[ktt(2019)]{kttf}
{K}eras to {T}ensor{F}low.
\newblock \url{https://github.com/amir-abdi/keras_to_tensorflow}, 2019.

\bibitem[sec(2019)]{securenncode}
{S}ecure{NN}: 3-{P}arty {S}ecure {C}omputation for {N}eural {N}etwork
  {T}raining.
\newblock \url{https://github.com/snwagh/securenn-public}, 2019.

\bibitem[Abadi et~al.(2016)Abadi, Agarwal, Barham, Brevdo, Chen, Citro,
  Corrado, Davis, Dean, Devin, Ghemawat, Goodfellow, Harp, Irving, Isard, Jia,
  J{\'{o}}zefowicz, Kaiser, Kudlur, Levenberg, Man{\'{e}}, Monga, Moore,
  Murray, Olah, Schuster, Shlens, Steiner, Sutskever, Talwar, Tucker,
  Vanhoucke, Vasudevan, Vi{\'{e}}gas, Vinyals, Warden, Wattenberg, Wicke, Yu,
  and Zheng]{tensorflow}
Mart{\'{\i}}n Abadi, Ashish Agarwal, Paul Barham, Eugene Brevdo, Zhifeng Chen,
  Craig Citro, Gregory~S. Corrado, Andy Davis, Jeffrey Dean, Matthieu Devin,
  Sanjay Ghemawat, Ian~J. Goodfellow, Andrew Harp, Geoffrey Irving, Michael
  Isard, Yangqing Jia, Rafal J{\'{o}}zefowicz, Lukasz Kaiser, Manjunath Kudlur,
  Josh Levenberg, Dan Man{\'{e}}, Rajat Monga, Sherry Moore, Derek~Gordon
  Murray, Chris Olah, Mike Schuster, Jonathon Shlens, Benoit Steiner, Ilya
  Sutskever, Kunal Talwar, Paul~A. Tucker, Vincent Vanhoucke, Vijay Vasudevan,
  Fernanda~B. Vi{\'{e}}gas, Oriol Vinyals, Pete Warden, Martin Wattenberg,
  Martin Wicke, Yuan Yu, and Xiaoqiang Zheng.
\newblock {T}ensor{F}low: {L}arge-{S}cale {M}achine {L}earning on
  {H}eterogeneous {D}istributed {S}ystems.
\newblock \emph{CoRR}, abs/1603.04467, 2016.
\newblock URL \url{https://arxiv.org/abs/1603.04467}.

\bibitem[Agrawal et~al.(2019)Agrawal, Shamsabadi, Kusner, and
  Gasc{\'{o}}n]{nitin}
Nitin Agrawal, Ali~Shahin Shamsabadi, Matt~J. Kusner, and Adri{\`{a}}
  Gasc{\'{o}}n.
\newblock {QUOTIENT:} {T}wo-{P}arty {S}ecure {N}eural {N}etwork {T}raining and
  {P}rediction.
\newblock In \emph{Proceedings of the 2019 {ACM} {SIGSAC} Conference on
  Computer and Communications Security, {CCS} 2019, London, UK, November 11-15,
  2019}, pages 1231--1247, 2019.

\bibitem[Beaver(1991)]{beaver}
Donald Beaver.
\newblock {E}fficient {M}ultiparty {P}rotocols {U}sing {C}ircuit
  {R}andomization.
\newblock In \emph{Advances in Cryptology - {CRYPTO} '91, 11th Annual
  International Cryptology Conference, Santa Barbara, California, USA, August
  11-15, 1991, Proceedings}, pages 420--432, 1991.

\bibitem[Boemer et~al.(2019)Boemer, Lao, Cammarota, and Wierzynski]{nhe}
Fabian Boemer, Yixing Lao, Rosario Cammarota, and Casimir Wierzynski.
\newblock {nGraph-HE}: {A} {G}raph {C}ompiler for {D}eep {L}earning on
  {H}omomorphically {E}ncrypted {D}ata.
\newblock In \emph{Proceedings of the 16th {ACM} International Conference on
  Computing Frontiers, {CF} 2019, Alghero, Italy, April 30 - May 2, 2019},
  pages 3--13, 2019.

\bibitem[Chandran et~al.(2019)Chandran, Gupta, Rastogi, Sharma, and
  Tripathi]{ezpc}
Nishanth Chandran, Divya Gupta, Aseem Rastogi, Rahul Sharma, and Shardul
  Tripathi.
\newblock {EzPC}: {Programmable and Efficient Secure Two-Party Computation for
  Machine Learning}.
\newblock In \emph{{IEEE} European Symposium on Security and Privacy,
  EuroS{\&}P 2019, Stockholm, Sweden, June 17-19, 2019}, pages 496--511, 2019.

\bibitem[Dahl et~al.(2018)Dahl, Mancuso, Dupis, Decoste, Giraud, Livingstone,
  Patriquin, and Uhma]{tfe}
Morten Dahl, Jason Mancuso, Yann Dupis, Ben Decoste, Morgan Giraud, Ian
  Livingstone, Justin Patriquin, and Gavin Uhma.
\newblock {Private Machine Learning in TensorFlow using Secure Computation}.
\newblock \emph{CoRR}, abs/1810.08130, 2018.
\newblock URL \url{http://arxiv.org/abs/1810.08130}.

\bibitem[Dalskov et~al.(2020)Dalskov, Escudero, and Keller]{quantizednn}
Anders P.~K. Dalskov, Daniel Escudero, and Marcel Keller.
\newblock Secure evaluation of quantized neural networks.
\newblock \emph{Proc. Priv. Enhancing Technol.}, 2020\penalty0 (4):\penalty0
  355--375, 2020.

\bibitem[Dathathri et~al.(2019)Dathathri, Saarikivi, Chen, Lauter, Maleki,
  Musuvathi, and Mytkowicz]{chet}
Roshan Dathathri, Olli Saarikivi, Hao Chen, Kristin Lauter, Saeed Maleki, Madan
  Musuvathi, and Todd Mytkowicz.
\newblock {CHET}: {An Optimizing Compiler for Fully-Homomorphic Neural-Network
  Inferencing}.
\newblock In \emph{Proceedings of the 40th {ACM} {SIGPLAN} Conference on
  Programming Language Design and Implementation, {PLDI} 2019, Phoenix, AZ,
  USA, June 22-26, 2019}, pages 142--156, 2019.

\bibitem[Demmler et~al.(2015)Demmler, Schneider, and Zohner]{aby}
Daniel Demmler, Thomas Schneider, and Michael Zohner.
\newblock {ABY} - {A Framework for Efficient Mixed-Protocol Secure Two-Party
  Computation}.
\newblock In \emph{22nd Annual Network and Distributed System Security
  Symposium, {NDSS} 2015, San Diego, California, USA, February 8-11, 2015},
  2015.

\bibitem[Goldreich et~al.(1987)Goldreich, Micali, and Wigderson]{gmw}
Oded Goldreich, Silvio Micali, and Avi Wigderson.
\newblock {How to Play any Mental Game or A Completeness Theorem for Protocols
  with Honest Majority}.
\newblock In \emph{Proceedings of the 19th Annual {ACM} Symposium on Theory of
  Computing, 1987, New York, New York, {USA}}, pages 218--229, 1987.

\bibitem[He et~al.(2016)He, Zhang, Ren, and Sun]{resnet}
Kaiming He, Xiangyu Zhang, Shaoqing Ren, and Jian Sun.
\newblock {Deep Residual Learning for Image Recognition}.
\newblock In \emph{2016 {IEEE} Conference on Computer Vision and Pattern
  Recognition, {CVPR} 2016, Las Vegas, NV, USA, June 27-30, 2016}, pages
  770--778, 2016.

\bibitem[Huang et~al.(2017)Huang, Liu, van~der Maaten, and
  Weinberger]{densenet}
Gao Huang, Zhuang Liu, Laurens van~der Maaten, and Kilian~Q. Weinberger.
\newblock {Densely Connected Convolutional Networks}.
\newblock In \emph{2017 {IEEE} Conference on Computer Vision and Pattern
  Recognition, {CVPR} 2017, Honolulu, HI, USA, July 21-26, 2017}, pages
  2261--2269, 2017.

\bibitem[Juvekar et~al.(2018)Juvekar, Vaikuntanathan, and
  Chandrakasan]{gazelle}
Chiraag Juvekar, Vinod Vaikuntanathan, and Anantha Chandrakasan.
\newblock {GAZELLE: A Low Latency Framework for Secure Neural Network
  Inference}.
\newblock In \emph{27th {USENIX} Security Symposium, {USENIX} Security 2018,
  Baltimore, MD, USA, August 15-17, 2018}, pages 1651--1669, 2018.

\bibitem[Kumar et~al.(2020)Kumar, Rathee, Chandran, Gupta, Rastogi, and
  Sharma]{cryptflow}
Nishant Kumar, Mayank Rathee, Nishanth Chandran, Divya Gupta, Aseem Rastogi,
  and Rahul Sharma.
\newblock Cryptflow: Secure tensorflow inference.
\newblock In \emph{2020 {IEEE} Symposium on Security and Privacy, {SP} 2020,
  San Francisco, CA, USA, May 18-21, 2020}, pages 336--353. {IEEE}, 2020.

\bibitem[Liu et~al.(2017)Liu, Juuti, Lu, and Asokan]{minionn}
Jian Liu, Mika Juuti, Yao Lu, and N.~Asokan.
\newblock {Oblivious Neural Network Predictions via MiniONN Transformations}.
\newblock In \emph{Proceedings of the 2017 {ACM} {SIGSAC} Conference on
  Computer and Communications Security, {CCS} 2017, Dallas, TX, USA, October 30
  - November 03, 2017}, pages 619--631, 2017.

\bibitem[Mishra et~al.(2020)Mishra, Lehmkuhl, Srinivasan, Zheng, and
  Popa]{delphi}
Pratyush Mishra, Ryan Lehmkuhl, Akshayaram Srinivasan, Wenting Zheng, and
  Raluca~Ada Popa.
\newblock {Delphi: A Cryptographic Inference Service for Neural Networks}.
\newblock In \emph{29th {USENIX} Security Symposium, {USENIX} Security 20},
  Boston, MA, 2020.

\bibitem[Mohassel and Rindal(2018)]{aby3}
Payman Mohassel and Peter Rindal.
\newblock {ABY}\({}^{\mbox{3}}\): {A Mixed Protocol Framework for Machine
  Learning}.
\newblock In \emph{Proceedings of the 2018 {ACM} {SIGSAC} Conference on
  Computer and Communications Security, {CCS} 2018, Toronto, ON, Canada,
  October 15-19, 2018}, pages 35--52, 2018.

\bibitem[Mohassel and Zhang(2017)]{secureml}
Payman Mohassel and Yupeng Zhang.
\newblock {SecureML: A System for Scalable Privacy-Preserving Machine
  Learning}.
\newblock In \emph{2017 {IEEE} Symposium on Security and Privacy, {S\&P} 2017,
  San Jose, CA, USA, May 22-26, 2017}, pages 19--38, 2017.

\bibitem[Nikolov et~al.(2018)Nikolov, Blackwell, Mendes, Fauw, Meyer, Hughes,
  Askham, Romera{-}Paredes, Karthikesalingam, Chu, Carnell, Boon, D'Souza,
  Moinuddin, Sullivan, Consortium, Montgomery, Rees, Sharma, Suleyman, Back,
  Ledsam, and Ronneberger]{stan}
Stanislav Nikolov, Sam Blackwell, Ruheena Mendes, Jeffrey~De Fauw, Clemens
  Meyer, C{\'{\i}}an Hughes, Harry Askham, Bernardino Romera{-}Paredes, Alan
  Karthikesalingam, Carlton Chu, Dawn Carnell, Cheng Boon, Derek D'Souza,
  Syed~Ali Moinuddin, Kevin Sullivan, DeepMind~Radiographer Consortium, Hugh
  Montgomery, Geraint Rees, Ricky Sharma, Mustafa Suleyman, Trevor Back,
  Joseph~R. Ledsam, and Olaf Ronneberger.
\newblock Deep learning to achieve clinically applicable segmentation of head
  and neck anatomy for radiotherapy.
\newblock \emph{CoRR}, abs/1809.04430, 2018.
\newblock URL \url{http://arxiv.org/abs/1809.04430}.

\bibitem[Oktay et~al.(2020)Oktay, Nanavati, Schwaighofer, Carter, Bristow,
  Tanno, Jena, Barnett, Noble, Rimmer, Glocker, O’Hara, Bishop,
  Alvarez-Valle, and Nori]{innereye}
Ozan Oktay, Jay Nanavati, Anton Schwaighofer, David Carter, Melissa Bristow,
  Ryutaro Tanno, Rajesh Jena, Gill Barnett, David Noble, Yvonne Rimmer, Ben
  Glocker, Kenton O’Hara, Christopher Bishop, Javier Alvarez-Valle, and
  Aditya Nori.
\newblock {Evaluation of Deep Learning to Augment Image-Guided Radiotherapy for
  Head and Neck and Prostate Cancers}.
\newblock \emph{JAMA Network Open}, 3\penalty0 (11):\penalty0
  e2027426--e2027426, 11 2020.
\newblock ISSN 2574-3805.
\newblock \doi{10.1001/jamanetworkopen.2020.27426}.
\newblock URL \url{https://doi.org/10.1001/jamanetworkopen.2020.27426}.

\bibitem[Pan et~al.(2016)Pan, Haffty, Falit, Buchholz, Wilson, Hahn, and
  Smith]{demand}
Hubert~Y Pan, Bruce~G Haffty, Benjamin~P Falit, Thomas~A Buchholz, Lynn~D
  Wilson, Stephen~M Hahn, and Benjamin~D Smith.
\newblock Supply and demand for radiation oncology in the united states:
  Updated projections for 2015 to 2025.
\newblock \emph{{Int J Radiat Oncol Biol Phys.}}, 96\penalty0 (3):\penalty0
  493--500, 2016.
\newblock \doi{10.1016/j.ijrobp.2016.02.064}.

\bibitem[Rathee et~al.(2020)Rathee, Rathee, Kumar, Chandran, Gupta, Rastogi,
  and Sharma]{cryptflow2}
Deevashwer Rathee, Mayank Rathee, Nishant Kumar, Nishanth Chandran, Divya
  Gupta, Aseem Rastogi, and Rahul Sharma.
\newblock Cryptflow2: Practical secure 2-party inference.
\newblock In \emph{Proceedings of the 2020 {ACM} {SIGSAC} Conference on
  Computer and Communications Security, {CCS} 2020}. {ACM}, 2020.

\bibitem[Riazi et~al.(2018)Riazi, Weinert, Tkachenko, Songhori, Schneider, and
  Koushanfar]{chameleon}
M.~Sadegh Riazi, Christian Weinert, Oleksandr Tkachenko, Ebrahim~M. Songhori,
  Thomas Schneider, and Farinaz Koushanfar.
\newblock {Chameleon: A Hybrid Secure Computation Framework for Machine
  Learning Applications}.
\newblock In \emph{Proceedings of the 2018 on Asia Conference on Computer and
  Communications Security, AsiaCCS 2018, Incheon, Republic of Korea, June
  04-08, 2018}, pages 707--721, 2018.
\newblock \doi{10.1145/3196494.3196522}.

\bibitem[Riazi et~al.(2019)Riazi, Samragh, Chen, Laine, Lauter, and
  Koushanfar]{xonn}
M.~Sadegh Riazi, Mohammad Samragh, Hao Chen, Kim Laine, Kristin~E. Lauter, and
  Farinaz Koushanfar.
\newblock {XONN: XNOR-based Oblivious Deep Neural Network Inference}.
\newblock In \emph{28th {USENIX} Security Symposium, {USENIX} Security 2019,
  Santa Clara, CA, USA, August 14-16, 2019}, pages 1501--1518, 2019.

\bibitem[Ronneberger et~al.(2015)Ronneberger, P.Fischer, and Brox]{unet}
O.~Ronneberger, P.Fischer, and T.~Brox.
\newblock U-net: Convolutional networks for biomedical image segmentation.
\newblock In \emph{Medical Image Computing and Computer-Assisted Intervention
  (MICCAI)}, volume 9351 of \emph{LNCS}, pages 234--241. Springer, 2015.
\newblock URL
  \url{http://lmb.informatik.uni-freiburg.de/Publications/2015/RFB15a}.
\newblock (available on arXiv:1505.04597 [cs.CV]).

\bibitem[Ryffel et~al.(2018)Ryffel, Trask, Dahl, Wagner, Mancuso, Rueckert, and
  Passerat{-}Palmbach]{pysyft}
Theo Ryffel, Andrew Trask, Morten Dahl, Bobby Wagner, Jason Mancuso, Daniel
  Rueckert, and Jonathan Passerat{-}Palmbach.
\newblock {A generic framework for privacy preserving deep learning}.
\newblock \emph{CoRR}, abs/1811.04017, 2018.
\newblock URL \url{http://arxiv.org/abs/1811.04017}.

\bibitem[Wagh et~al.(2019)Wagh, Gupta, and Chandran]{securenn}
Sameer Wagh, Divya Gupta, and Nishanth Chandran.
\newblock {SecureNN: 3-Party Secure Computation for Neural Network Training}.
\newblock \emph{PoPETs}, 2019\penalty0 (3):\penalty0 26--49, 2019.

\bibitem[Wang et~al.(2020)Wang, Nie, Qu, Shao, Lian, Wang, and Shen]{shuai}
Shuai Wang, Dong Nie, Liangqiong Qu, Yeqin Shao, Jun Lian, Qian Wang, and
  Dinggang Shen.
\newblock {CT} male pelvic organ segmentation via hybrid loss network with
  incomplete annotation.
\newblock \emph{{IEEE} Trans. Medical Imaging}, 39\penalty0 (6):\penalty0
  2151--2162, 2020.
\newblock \doi{10.1109/TMI.2020.2966389}.
\newblock URL \url{https://doi.org/10.1109/TMI.2020.2966389}.

\bibitem[Yao(1986)]{yao}
Andrew~Chi{-}Chih Yao.
\newblock {How to Generate and Exchange Secrets (Extended Abstract)}.
\newblock In \emph{27th Annual Symposium on Foundations of Computer Science,
  Toronto, Canada, 27-29 October 1986}, pages 162--167. {IEEE} Computer
  Society, 1986.
\newblock \doi{10.1109/SFCS.1986.25}.

\bibitem[Zhu et~al.(2018)Zhu, Iordanescu, Karmanov, and
  Zawaideh]{chestxray2018}
Xiaoyong Zhu, George Iordanescu, Ilia Karmanov, and Mazen Zawaideh, Mar 2018.
\newblock URL
  \url{https://blogs.technet.microsoft.com/machinelearning/2018/03/07/using-microsoft-ai-to-build-a-lung-disease-prediction-model-using-chest-x-ray-images/}.

\end{thebibliography}
\endgroup

\end{document}